\documentclass[journal]{IEEEtran}

\usepackage{graphicx}
\usepackage{cite}
\usepackage{xfrac}
\usepackage{amsmath}
\usepackage{mathtools}
\usepackage{xcolor}

\begin{document}

%%%%%%%%%%%%%%%%%%%%%%%%%%%%%%%%%%%%%%%%%%%%%%%%%%%%%%%%%%%%%%%%%%%%%%%%%%
%%%%%%%%%%%%%%%%%%%% TITLE %%%%%%%%%%%%%%%%%%%%%%%%%%%%%%%%%%%%%%%%%%%%%%%
\title{Automated Fovea Detection Based on Unsupervised Retinal Vessel Segmentation Method}
\author{Meysam Tavakoli,
		Patrick Kelley,
		Mahdieh Nazar,
        Faraz Kalantari % <-this % stops a space
%\thanks{} to gain access to the first footnote area
\thanks{M. Tavakoli and P. Kelley are with the Dept. of Physics, Indiana University-Purdue University, Indianapolis, IN, USA.}% <-this % stops a space
\thanks{M. Nazar is with the Biomedical Sciences Department, Shahid Beheshti University of Medical Sciences, Tehran, Iran}% <-this % stops a space
\thanks{F. Kalantari is with Department of Radiation Oncology, University of Texas Southwestern Medical Center, Dallas, TX, USA
}% <-this % stops a space
\\
\textbf{To appear in: 2017 IEEE Nuclear Science Symposium and Medical Imaging Conference (NSS/MIC)\\
DOI: 10.1109/NSSMIC.2017.8533061}
}
\maketitle

%%%%%%%%%%%%%%%%%%%%%%%%%%%%%%%%%%%%%%%%%%%%%%%%%%%%%%%%%%%%%%%%%%%%%%%%%%
%%%%%%%%%%%%%%%%%%%% ABSTRACT %%%%%%%%%%%%%%%%%%%%%%%%%%%%%%%%%%%%%%%%%%%%
%%%%%%%%%%%%%%%%%%%%%%%%%%%%%%%%%%%%%%%%%%%%%%%%%%%%%%%%%%%%%%%%%%%%%%%%%%
\begin{abstract}
The Computer Assisted Diagnosis systems could save workloads and give objective diagnostic to ophthalmologists. At first level of automated screening of systems feature extraction is the fundamental step. One of these retinal features is the fovea. The fovea is a small fossa on the fundus, which is represented by a deep-red or red-brown color in color retinal images. By observing retinal images, it appears that the main vessels diverge from the optic nerve head and follow a specific course that can be geometrically modeled as a parabola, with a common vertex inside the optic nerve head and the fovea located along the apex of this parabola curve. Therefore, based on this assumption, the main retinal blood vessels are segmented and fitted to a parabolic model. With respect to the core vascular structure, we can thus detect fovea in the fundus images. For the vessel segmentation, our algorithm addresses the image locally where homogeneity of features is more likely to occur. The algorithm is composed of 4 steps: multi-overlapping windows, local Radon transform, vessel validation, and parabolic fitting. In order to extract blood vessels, sub-vessels should be extracted in local windows. The high contrast between blood vessels and image background in the images cause the vessels to be associated with peaks in the Radon space. The largest vessels, using a high threshold of the Radon transform, determines the main course or overall configuration of the blood vessels which when fitted to a parabola, leads to the future localization of the fovea. In effect, with an accurate fit, the fovea normally lies along the slope joining the vertex and the focus. The darkest region along this line is the indicative of the fovea. To evaluate our method, we used 220 fundus images from a rural databse (MUMS-DB) and one public one (DRIVE).
The results show that, among 20 images of the first public database (DRIVE) we detected fovea in 85\% of them. Also for the MUMS-DB database among 200 images we detect fovea correctly in 83\% on them.

\end{abstract}


% \fi
\IEEEpeerreviewmaketitle

%%%%%%%%%%%%%%%%%%%%%%%%%%%%%%%%%%%%%%%%%%%%%%%%%%%%%%%%%%%%%%%%%%%%%%%%%%
%%%%%%%%%%%%%%%%%%%% Introduction %%%%%%%%%%%%%%%%%%%%%%%%%%%%%%%%%%%%%%%%
%%%%%%%%%%%%%%%%%%%%%%%%%%%%%%%%%%%%%%%%%%%%%%%%%%%%%%%%%%%%%%%%%%%%%%%%%%
\section{Introduction}
\IEEEPARstart{T}{he} Retinal images are widely used for screening of systemic diseases such as diabetic retinopathy (DR). Early detection and treatment of these diseases are critical to avoid vision loss and blindness. In the old fashion way of diagnosis, the ophthalmologists do all evaluations on retinal images to give a diagnostic result. Computer Assisted Diagnosis (CAD) systems could save workloads and may be used as an objective diagnostic tool for ophthalmologists  \cite{Khansari-TMI, TavakoliFA, Tavakoli-SPECT}. 

Diabetes is a complicated disease that influences the blood vessel network throughout
the body especially in the eyes. The result of this affection of diabetes on retinal blood
vessels is DR. In the united state DR is the leading cause of blindness between adults \cite{Khansari-DR} and in the western country \cite{Englmeier}. In DR the microvasculature network changes and
one method for treating that, is laser surgery therapy if this change detected early as
soon as possible. It effect pathological changes in the retinal vessel network such as
MAs, HEs, exudates as well as intraretinal microvascular abnormalities, venous
beading and neovascularisations. According to \cite{Hoover1}, other than of DR, two
circumstances that are diagnosed from the manifestation of blood vessels are
Hypertension and arteriosclerosis. As well as field of bioinformatics is another
application of blood vessel detection \cite{Xu ZW}.
To modify retinal vasculature there are two methods: using opthalmoscopy and by
analyzing fundus photographs.
Detection of retinal vasculature network especially abnormal vessel was done by
ophthalmologists, which is time consuming work and is associated with error and
fatigue. As well as using non-professional health workers reduce our identification
efficiency by 50\% \cite{Cornforth}.
In other to increase the evaluation of the retinal images image processing techniques
are required to extract suitable imformation about alteration of retinal vessel tree.
Automated blood vessel segmentation is critical step in screening of some diseases
like that: DR \cite{Sinthanayothin2} assessment of the retinopathy of prematurity \cite{Heneghan}, or for determining
of optic nerve head \cite{Hoover2} and fovea cite{Welfer}.

Our CAD approach is to use vessel segmentation and parabola fitting to find the fovea. The fovea is the spot with the sharpest and clearest vision because it has no vessels blocking this dense center of cones, the photoreceptors that process bright light and color. So if disease or lesions sets in, the fovea can be a sensitive area that can quickly lead to severe damage to the patients’ overall vision \cite{Mehdizadeh-color}. And because the fovea is the smallest part of the macula region and this area doesn’t have a well-defined structure, the fovea is a rather more difficult part of the retina to detect than the optic nerve head (ONH).

 Vessel segmentation is the a corner-stone in detection of other retinal land marks such as fovea \cite{Niemeijer1, sinthanayothin}. Moreover, it is critical step in screening of diabetic retinopathy (DR)\cite{Mehdizadeh-color, Gagnon L, Sherwani SM, Muangnak N, Constante P, Khansari-DR2}.\\
Before fovea and vessel detetction, the fundus image has to be preprocessed to ensure adequate level of success in detection. Here we applied two different methods top-hat transform. \\
There are several algorithms for retinal vessel segmentation \cite{sinthanayothin, Heneghan, Sinthanayothin2, Hoover1, Staal, Soares, Jiang X, Wu D, Tolias, Zana, Walter, Tavakoli-ONH2}. 

A model or template acquired from a set of ideal training retinal set was first used for detecting the fovea. One of the earlier methods to localize the fovea with templates was Sinthanayothin et al. \cite{sinthanayothin}. They used a template of typical fovea intensities to correlate with the intensities of the subimages. The template of the typical fovea was defined to follow a Gaussian distribution with a fixed standard deviation for the pixel darkness with a total size of 40×40 pixels. The subimages were then compared to the template to compute a correlation coefficients and after passing a threshold, the candidate regions were selected. The candidate with the darkest center that matched the most plausible distance from the previously detected optic disk was ultimately determined to be the detected fovea. Singh et al \cite{Singh} used a form of landmark of the retinal image that steers the program to localize the optic disk and fovea. They enhanced the contrast of the retinal image and used an intensity profile of a dark image template to compare and discern the fovea.
Another method that obtained good results in detecting the fovea were in using morphological operations. Sekhar et al. \cite{Sekhar} used morphological operations to detect the optic disk and then Hough Transform to detect the boundary of the optic disk. The detected optic disk diameter gives a region of interest (ROI) on where the macula and fovea should be. After some thresholding and morphological opening operations, the fovea center is ultimately obtained. Welfer similarly used morphological operations to identify the optic disk, the optic disk boundary and thus the diameter, then find a ROI where morphological operations and constraints were applied to the ROI until the fovea region was selected and detected. 
Some methods that inspired the development of our proposed methods stems from vessel identification, point distribution model and template features of the fovea. This approach by Fleming et al. \cite{Fleming} was used to detect the main vessels called the temporal arcade and then to use semi-circles as a template guide to detect the optic disk and fovea. Niemeijer et al. \cite{Niemeijer1} used a combination of template-based feature extraction and feature regression analysis. Relying on a circular template trained from 500 images, the optic disk is compared to the template based on the image’s point distribution model of the intensity features and vascular features such as vessel thickness and orientation. A k-Nearest Neighbor (kNN) regressor on these features is used to predict the distance to the object of interest. Upon finding the optic disk, the fovea is then inferred by the optic disk location and detected from this search parameter. Another method, that Li and Chutatape \cite{Chutatape} used, most closely matches with our proposed method. They used principal component analysis (PCA) which is a top-down strategy of finding common optic disk features from a training set. The PCA is comprised of three steps: obtaining a disk shape from highest 1\% grayscale intensity levels, projection of the image’s disk shape on the mean disk space with corresponding statistical variances from a training set, and calculating the disk space to obtain the optic disk. The disk boundary is constructed from a modified active shape model (ASM) which uses 48 landmark points to compare to a point distribution model (PDM) that was learned from a training set and fits to the landmark points in the image. The main course of the vessel is also obtained from landmark points that are used to fit the parabolic curve that sets a searching area to find the fovea. 
All these methods rely on determining a feature that is common between the fovea, the optic disk and/or the overall retina. Like Li and Chutatape \cite{Chutatape}, we use the feature of the vessels assuming a parabolic shape to lead us to detect the fovea. The difference is in the way we segment and utilize the retinal vessels. In the following sections, we outline the images we used in the Materials section. The Methodology section describes the preprocessing and vessel segmentation steps that use Radon Transform on sub-images which are recombined to create a vessel map. The thickest vessels create a vessel landmark points which we fit a parabolic curve. The vertex is assumed to be the fovea center. We report our outcome in the Results section and examine and consider the limitations and future work in the Discussion section. \\
\begin{table}[h!]
%% increase table row spacing, adjust to taste
\renewcommand{\arraystretch}{1.3}
% if using array.sty, it might be a good idea to tweak the value of
% \extrarowheight as needed to properly center the text within the cells
\caption{COMPARISON THE RESULTS OF SOME IMPORTANT METHODS IN FOVEA DETECTION}
\label{table1}
\centering
% Some packages, such as MDW tools, offer better commands for making tables
% than the plain LaTeX2e tabular which is used here.
\resizebox{\columnwidth}{!} {
\begin{tabular}{|c|c c|c c|}
\hline
%\rowcolor{lightgray}

Study & Method & No. of Images & Accuracy & year\\
\hline\hline
Sinthanayothin et al. \cite{sinthanayothin} & NN Algorithm
(Template-Based Method) & 112 & 80.4\% & 1999\\
\hline
Li and Chutatape \cite{Chutatape} & Modified Active Shape Model
(Parabola fitting) & 35 & 100\% & 2004 \\
\hline 
Fleming et al. \cite{Fleming} & Main Vessel detection with Semiellipse fitting & 1056 & 96.6\% & 2007 \\
\hline
Tobin et al. \cite{Tobin} & ONH Detection and Macula Localization & 345 & 92.5\% & 2007 \\
\hline
Singh et al. \cite{Singh}& Appearance-Based Method & 502 & 96.61\% & 2008 \\
\hline
Sekhar et al. \cite{Sekhar}& Hough transform and morphological operators & 44 & 100\% & 2008 \\
\hline
Niemeijer et al. \cite{Niemeijer1} & Point Distribution Model and kNN Regression & 500 & 96.80\% & 2009 \\
\hline
Welfer et al. \cite{Welfer} & Selection of ROI and morphology & 126 & 96.1\% & 2011 \\
\hline
\end{tabular}
}
\end{table}

%%%%%%%%%%%%%%%%%%%%%%%%%%%%%%%%%%%%%%%%%%%%%%%%%%%%%%%%%%%%%%%%%%%%%%%%%%
%%%%%%%%%%%%%%%%%%%% Proposed Methodology %%%%%%%%%%%%%%%%%%%%%%%%%%%%%%%%
%%%%%%%%%%%%%%%%%%%%%%%%%%%%%%%%%%%%%%%%%%%%%%%%%%%%%%%%%%%%%%%%%%%%%%%%%%
\section{Method}

%    h (here) - same location
%    t (top) - top of page
%    b (bottom) - bottom of page
%    p (page) - on an extra page
%    ! (override) - will force the specified location

\subsection{Materials}

For the fovea detection, two databsets, one rural and one publicly available databases, were used. The first rural database was named  Mashhad University Medical Science Database (MUMS-DB). A total of 220 color images were captured which 200 were with DR and 20 normal from both left and right eyes of patientsl. The train set consists of 20 images and our test set consists of 200 images (180 images with DR and 20 images without DR). Images were taken with
a field of view (FOV) of 50° under the same lighting conditions. According to the National
Screening Committee standards all the images are captured by using a TOPCON
(TRC-50EX), 50° Mydriatic retinal camera.
The acquired image resolution is $2896\times1944$ TIFF format  \cite{tavakoli-fluoresceinangiography, Tavakoli-ONH1}. The second database, DRIVE, consisting of 40 images with image resolution of $768\times584$ pixels in which 33 cases did not have any sign of DR and 7 ones showed signs of early or mild DR with a 45° FOV. This database is divided into two sets: testing and training set, each of them containing 20 images \cite{researchsection}.

\subsection{Preprocessing}

 The preprocessing step provides us an image with high possible vessel and background contrast and also unifies the histogram of the images. Here we applied top-hat transform \cite{tavakoli-radon} for preprocessing step \\
 Although retinal images have three components (R, G, B), their green channel has the best contrast between vessel and background; so the green channel is selected as input image.  
Here, we also used mathematical morphology operators.  Morphological operations work with two parts. The first one is the image to be processed and the second is called structure element. 
Erosion is used to reduce the objects in the image with the structure element, also known as the kernel. In contrast, dilation is used to increase the objects in the image. Secondary operations that depend on erosion and dilation are opening and closing operations. Opening, denoted as $f \circ b$, is applying an erosion followed by a dilation operation. The b represents the structure element. On the other hand, closing is first applying dilation then erosion. It is denoted as $f \bullet b$. Building from opening and closing operations, the top-hat transform is defined as the difference between the input image and its opening or closing.
The top-hat transform is one of the important morphological operators \cite{tavakoli-twopreprocessingsteps}. Based on dilation and erosion, opening and closing denoted by $f \circ b$ and $f \bullet b$ are defined. The top-hat transform is defined as the difference between the input image and its opening. The top-hat transform includes white top-hat transform (WTH) and black top-hat transform (BTH) are defined by:

\begin{equation} \label{eq:sensitivity}
\begin{cases}
WTH(x,y)= f(x,y) - f \circ b(x,y) \\
BTH(x,y)= f \bullet b(x,y) - f(x,y) \\
\end{cases}
\end{equation}

\begin{figure}[h!]
\includegraphics[width=\linewidth]{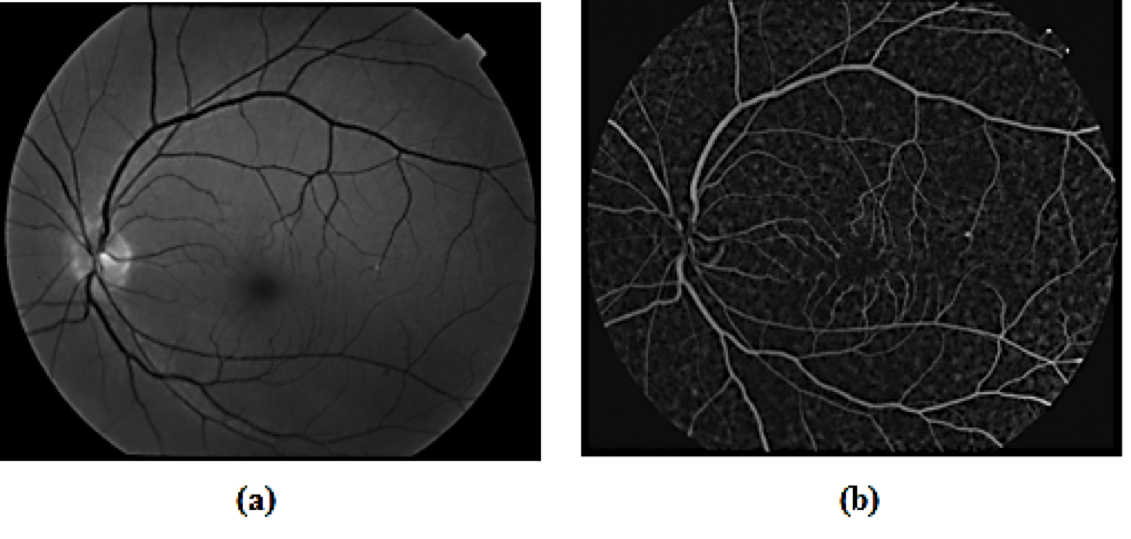}
\centering
\caption{(a) Fundus image from MUMS-DB (b) top-hat result and contrast stretching (c) result of subtraction of top-hat and filtered top-hat image.}
\label{fig2}
\end{figure}

In our preprocessing the basic idea is increasing the contrast between the vessels and background regions of the image. WTH or BTH extract bright and dim image regions corresponding to the used structure element. So, using the concept of WTH or BTH is one way of image enhancement through contrast enlarging based on top-hat transform.
In the fundus images, the background brightness is not the same in the whole image. This background variation would lead to missed vessels or false vessel detection in the following steps. Moreover, in I, background is brighter than the details, however the vessels and other components are preferred to appear brighter than background. To deal with the problem, I is inverted as shown in I= 255 - I.
Since we need a uniform background, to decrease the intensity variations in vessels background, we were firstly applied WTH(x,y) on image. It gave a high degree of differentiation between these features and background. A top-hat transformation was based on a ‘‘disk structure element’’ whose diameter was empirically found that the best compromise between the features and background. The disk diameter depended on the input image resolution.
After top-hat transformation, we used contrast stretching to change the contrast or brightness of an image. The result was a linear mapping of a subset of pixel values to the entire range of grays, from the black to the white, producing an image with much higher contrast. Filtering a region is the process of applying a filter to a region of interest in an image,
where a binary mask defines the region.
In this situation we used an averaging filter on the result of image from last section
(top-hat result) after that, we subtract the image from last section with the result of
applying averaging filter. Before this section, in image result from top-hat transform,
there are some variations in the image and some points like noise that without
eliminate these points maybe supposed as some part of vessels, that increase false
positive rate of our algorithm and after applying filter and subtraction this variation was removed. In other words, this section is applied for better removing background
variation for better detecting of vessels \cite{tavakoli-fluoresceinangiography}. The result of first step is shown in Fig. \ref{fig2}. 

\subsection{Main Procedure}
Feature extraction is a crucial step for automated screening of systems. One of these important retinal features is the fovea. Therefore, in this work we are looking for the fovea localization. The fovea is a small fossa on the fundus, which is represented by a deep-red or red-brown color in color retinal images. It is the darkest part in most of the fundus images. Its geometrical relation to other retinal landmarks is employed to locate the fovea robustly.\\
To do this we need first to segment the blood vessels.
Blood vessels can be described as bright curvilinear objects opposite of a darker
background with unclear edges.
The retinal blood vessels are non-uniform in intensity, length and width throughout
the image. Our approach  addresses the image locally and regionally where homogeneity of the vessel is more likely to happen. The algorithm is composed of 3 steps: Generation of sub-images (multi-overlapping windows), vessel segmentation, and fovea detection. 
In order to extract fovea, it should be extracted in local windows.
\subsubsection{Multi-overlapping window}
In the proposed algorithm, each image was partitioned into overlapping widows in the first step. To find objects on
border of sub-images, an overlapping pattern of sliding windows was defined. For determining the size of each sub-image or
sliding window our knowledge database was used. In this regard, minimum and maximum sizes of targeted object specify the
size of the windows (n) \cite{tavakoli-radon}.
The $\textit{n}$ has a direct effect on the extraction accuracy. 

\begin{figure}[h!]
\includegraphics[width=2in]{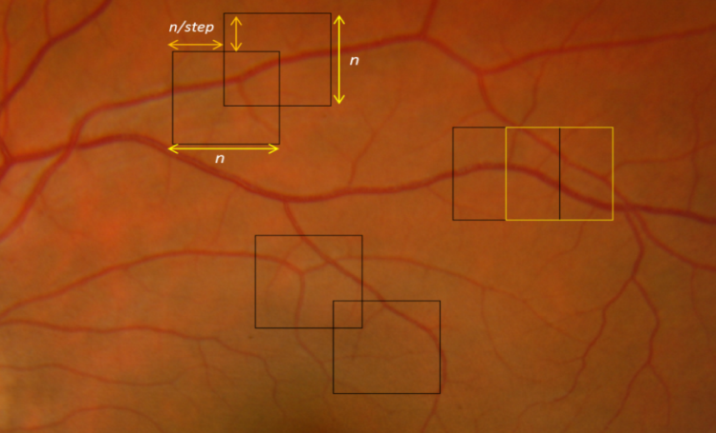}
\centering
\caption{Window size example.}
\label{fig3}
\end{figure}

\subsubsection{Vessel Segmentation}
In this study a Radon transform-based algorithm is proposed for detection of vessel
tree in fundal images.
The Radon transform is less sensitive to noise in the image, because the intensity
fluctuations due to noise tend to be cancelled out by the process of integration \cite{Choi}
The Radon transform is able to transform line-containing images into a domain where
each line in the image gives a peak or a valley in that domain.
In spite of morphologic operators which are image based and couldn't provide
numerical data directly, by using radon transform object finding and description will
be done simultaneously.
For line detection some numerical data such as line width, and line angle could be
achieved directly from Radon matrix.
\\
In the proposed algorithm, Green (G) component of sub-images are selected for
applying local Radon transform (RT). The reason is the high contrast between blood
vessels and image background in G component. G component is then complemented;
this makes the vessels to be associated with peaks in Radon space (not valleys). In
spite of the high contrast, we enhance the quality of sub-images before applying RT.
Image enhancement improves the sub-image contrast and causes more discrimination
between sub-vessels and retinal background and increases the total accuracy.
The RT of a function f(x,y), denoted as g(s,$\theta$), is defined as its line integral projected
at an angle $\theta$ from the y-axis, at a distance s from the origin. The RT, which is a
parameter transform, is defined as:

\begin{equation}
g_{\theta}(s) = \int_{0}^{y}\int_{0}^{x} f(x,y)\delta(s-xcos\theta-ysin\theta) dx dy 
\end{equation}
where\\
- g(s,$\theta$) is the one-dimensional projection of f(x,y) at an angle $\theta$.\\
- f(x,y) is the image intensity at (x,y). \\
- $\delta$ is the Dirac delta function. \\
- s = x cos$\theta$ + y sin$\theta$ , is the distance from the origin to the line being integrated.\\
The peak and the corresponding projection angle are detected in the  Radon
matrix. The profile in which peak occurs, is a candidate that might contain a sub-
vessel. This profile is further analyzed for validation of candidate sub-vessel.

\begin{figure}[h!]
\includegraphics[width=\linewidth]{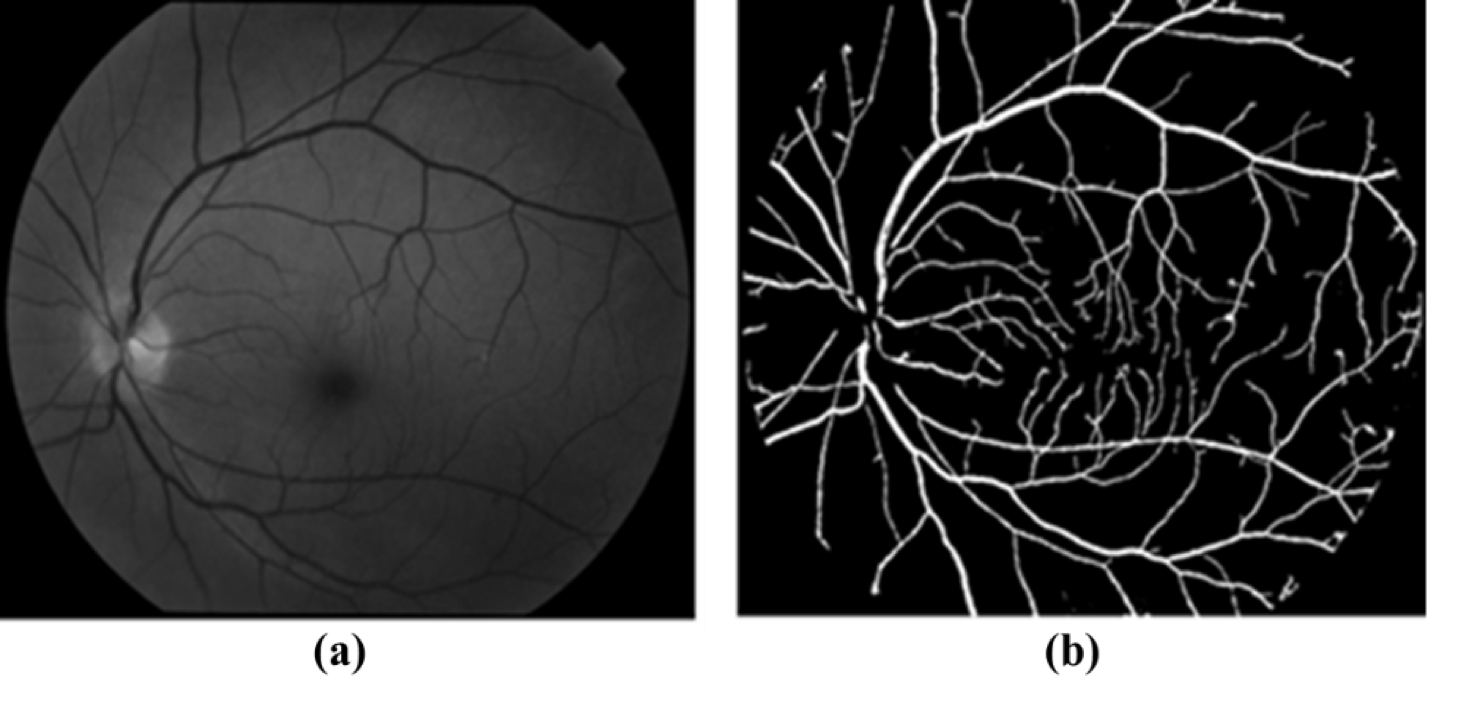}
\centering
\caption{(a) Original Green channel fundus image from local database (MUMS-DB), (b) result of segmentation using RT.}
\label{fig33}
\end{figure}

\begin{figure}[h!]
\includegraphics[width=\linewidth]{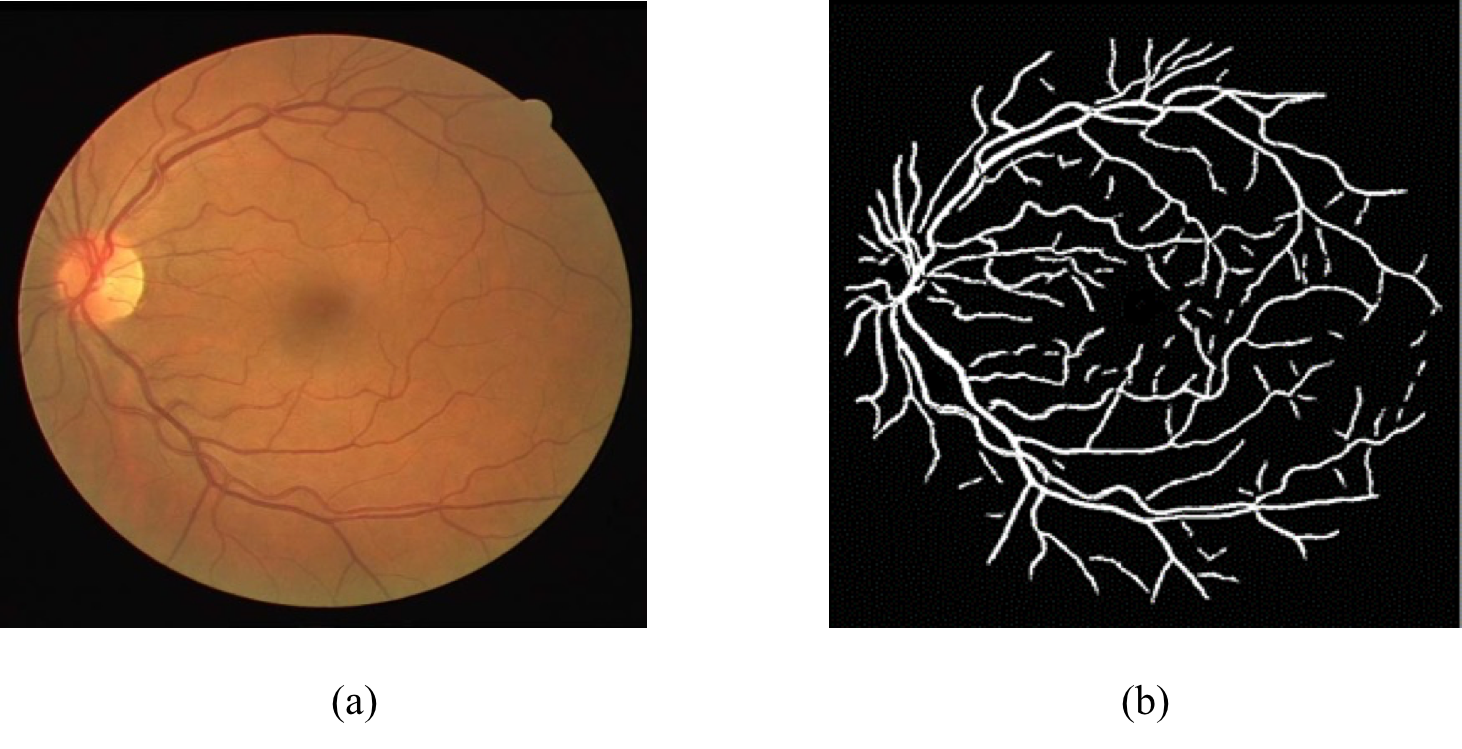}
\centering
\caption{(a) Original fundus image from DRIVE, (b) result of segmentation using RT.}
\label{fig55}
\end{figure}
\subsubsection{Fovea Detection}
It is apparent in retinal images that the main vessels diverge from the ONH and follow a specific course that can often be geometrically modeled as parabola, with a common vertex inside the ONH and the fovea is apex of this parabola curve. Based on this assumption, the main retinal blood vessels are segmented and fitted with a parabolic model. We can thus detect the fovea in the fundus images with respect to the entire vascular structure. Here we assume a Cartesian coordinate system, so this parabolic curve can be described as \cite{Foracchia}:
\begin{equation}
x = \lbrace (x,y): x=ay^2\rbrace
\end{equation}
Where \textit{a} is the focus or the amount and direction of opening of the parabola. 
\begin{figure}[h!]
\includegraphics[width=\linewidth]{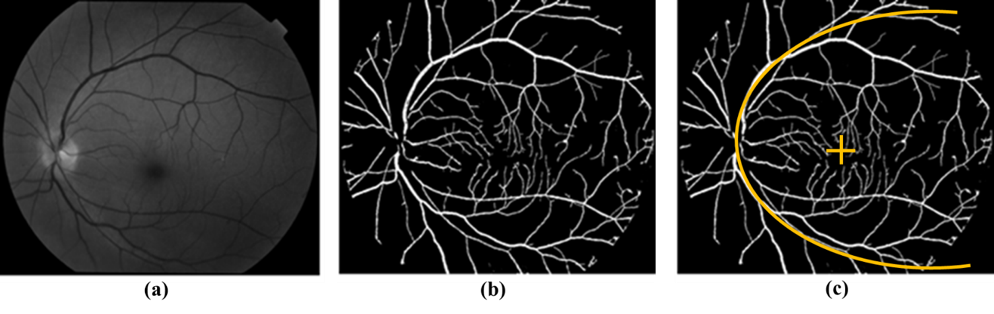}
\centering
\caption{(a) Original Green channel fundus image from local database (MUMS-DB), (b) result of segmentation using RT, (c) result of Fovea detection.}
\label{fig44}
\end{figure}
\begin{figure}[h!]
\includegraphics[width=\linewidth]{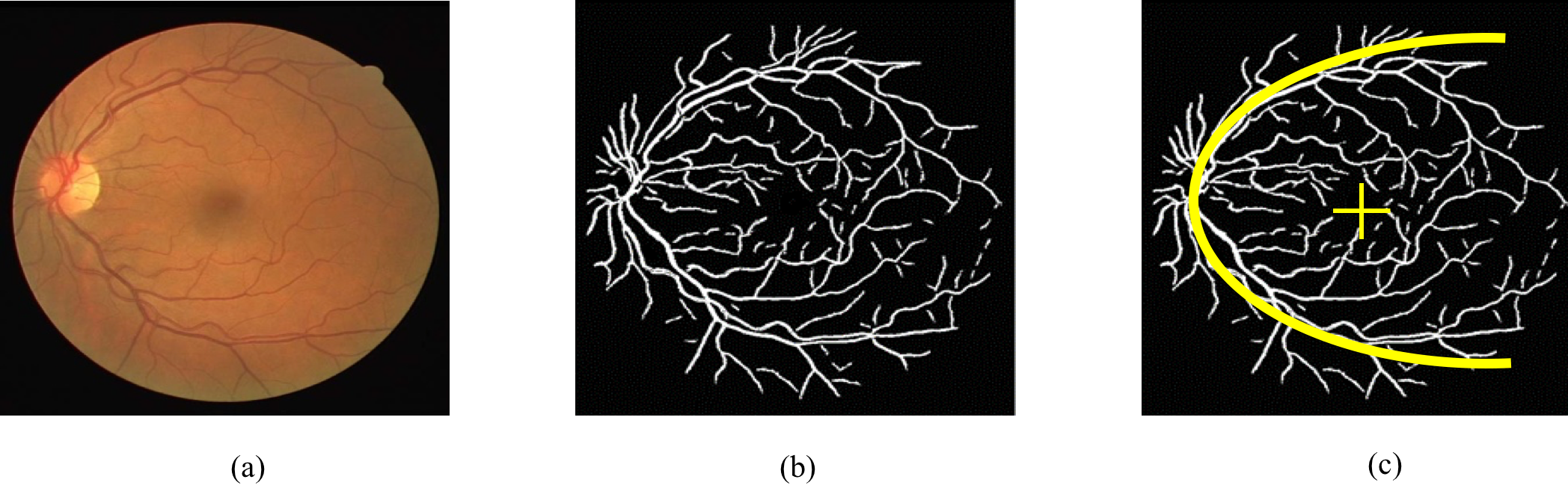}
\centering
\caption{(a) Original Green channel fundus image from DRIVE (b) result of segmentation using RT, (c) result of Fovea detection.}
\label{fig555}
\end{figure}
%%%%%%%%%%%%%%%%%%%%%%%%%%%%%%%%%%%%%%%%%%%%%%%%%%%%%%%%%%%%%%%%%%%%%%%%%%
%%%%%%%%%%%%%%%%%%%% Experimental Results %%%%%%%%%%%%%%%%%%%%%%%%%%%%%%%%
%%%%%%%%%%%%%%%%%%%%%%%%%%%%%%%%%%%%%%%%%%%%%%%%%%%%%%%%%%%%%%%%%%%%%%%%%%
\section{Experimental Results}

To calculate the efficiency of the current method in fovea detection and also to compare the results with other reported studies, it is necessary to compare all pixels of the final automated segmentated images with the manual segmentation or grand truth (GT) files. For the evaluation, we used the concept of sensitivity and specificity.
The results for the automated method compared to the GS were calculated for each image. The higher the sensitivity and specificity values, the better the procedure. Regarding these calculations the proper metrics are defined as \cite{tavakoli-fluoresceinangiography}: 

\begin{equation} \label{eq:sensitivity}
\begin{cases}
Sensitivity = \frac{TP}{TP+FN} \\
Specificity = \frac{TN}{TN+FP} \\
\end{cases}
\end{equation}

Where TP is true positive, TN is true negative, FP is false positive and FN is false negative.

\subsection{Training and Test Set for the Image Database}
\begin{figure}[h!]
\includegraphics[width=\linewidth]{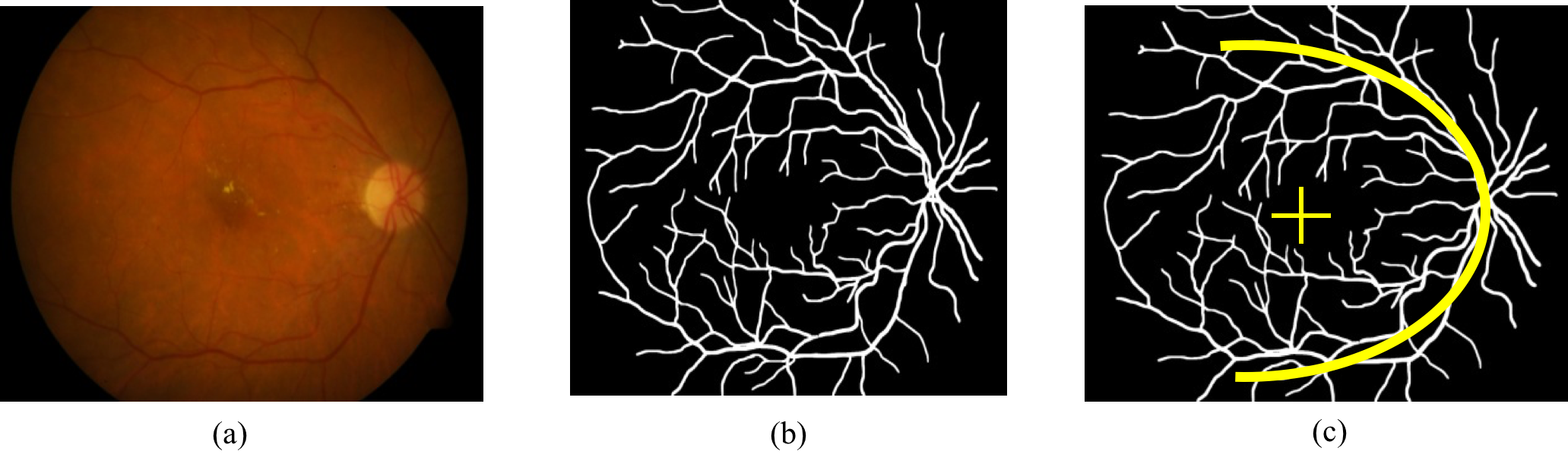}
\centering
\caption{(a) Original Green channel fundus image (b) result of segmentation using RT, (c) result of Fovea detection.}
\label{fig66}
\end{figure}

In this study, we used 40 images for a training set which was for learning purpose and to set our parameters in the algorithm. This consisted of 20 images from all databases (MUMS-DB,  DRIVE). The test set for the test purposes and to evaluate the method consisted of 220 fundus images of which 200 for MUMS-DB, and 20 for the DRIVE. After setting up the parameters of our algorithm using training set, the methods were tested in each image of the databases in test set.  Some results are shown in Fig. \ref{fig44},  Fig. \ref{fig555}, and Fig. \ref{fig66} from MUMS-DB and DRIVE.

The sensitivity of the threshold was also characterized along the Equation (8). All of three databases, the value $Th$ is varied over the range [0, 5].  Parameters used in the algorithm are shown in Table \ref{table2}.

\begin{table}[h!]
%% increase table row spacing, adjust to taste
\renewcommand{\arraystretch}{1.3}
% if using array.sty, it might be a good idea to tweak the value of
% \extrarowheight as needed to properly center the text within the cells
\caption{PARAMETERS USED IN OUR APPROACH FOR ALL THE THREE  DATABASES}
\label{table2}
\centering
% Some packages, such as MDW tools, offer better commands for making tables
% than the plain LaTeX2e tabular which is used here.
\resizebox{\columnwidth}{!} {
\begin{tabular}{|c|c|c|c|c|c|}
\hline
%\rowcolor{lightgray} 
Database & No. of Images & Window Size (n) & Step & Th \\
\hline\hline
MUMS-DB & 100 & 62 & 5 & [0,5] \\
\hline
DRIVE & 20 & 15 & 6 & [0,5] \\
\hline 
MESSIDOR & 100 & 40 & 5 & [0,5] \\
\hline
\end{tabular}
}
\end{table}

Statistical information about the sensitivity and specificity measures is extracted. The higher the sensitivity and specificity values, the better the procedure. For all retinal images of test set (220 images), our reader labeled the fovea on the images and the result of this manual detections were saved to be analyzed further. 

The result show that, among 40 images of the DRIVE database we detected fovea in 85\% of them. For the MUMS-DB database among 200 images we detect fovea correctly in 83\% of them. 
 \\

%%%%%%%%%%%%%%%%%%%%%%%%%%%%%%%%%%%%%%%%%%%%%%%%%%%%%%%%%%%%%%%%%%%%%%%%%%
%%%%%%%%%%%%%%%%%% Discussion and Conclusion %%%%%%%%%%%%%%%%%%%%%%%%%%%%%
%%%%%%%%%%%%%%%%%%%%%%%%%%%%%%%%%%%%%%%%%%%%%%%%%%%%%%%%%%%%%%%%%%%%%%%%%%
\section{Discussion and Conclusion}

Since fundus images are nowadays in the digital format, it is possible to create a
computer-based system that automatically detects landmarks from fundus images \cite{welikala, welikala2, Khansari-optics}. An
automatic system would save the working time of well-paid clinicians, and letting
hospitals to use their valuable resources in other important tasks. It could also be possible to check more people and more often with the help of an automatic
screening system, since it would be more inexpensive than screening by humans.
In this study, We present an automatic method for detecting the fovea, that is used for estimating the diabetic macular edema severity levels within the context of this project. Our method identifies the fovea center as a point (a pixel) and has been validated based on two public databases of retinal images. The accuracy of the fovea detection refers to how close is the automatically detected fovea center point (i.e. the distance in pixels) to the fovea center hand-labeled by an expert. 
Our approach uses a vessel segmentation and reconstruction technique \cite{tavakoli-radon} that relies on morphological operations and Radon Transform and a characteristic trait of vessels to predict the location of the fovea. Unlike other works, we make no template from a training set and only use one general feature, parabolic fitting, to detect the fovea. 
There are however current limitations. We first require a more robust fitting model, one that will need to rotate the image in the instances that the vessels arc neither purely up nor perfectly sideways. Another study needs to be performed to ascertain whether parabolic or higher order fitting is relevant to different or diseased retinas. The second limitation is to improve on the generalization of the location of the fovea when not directly lying on the vertex of the parabola. Future work is to create a line from the apex of the parabola through the vertex and find a region of possible candidate locations for the fovea.  This would provide a more comprehensive way to find and detect the fovea. Our proposed algorithm combines many of the powerful techniques from previous works and would be a more generic computational approach that doesn’t require previous exemplary images to base a template on. 

One of the most important parts of in this study is to use multi-overlapping window to
prevent error caused by global techniques.\\
One advantage of the automatic screening system is that it is deterministic, in other
words, it always classifies funduses in the similar way. There will always exist
differences in the backgrounds and education of human screeners, which causes
dispersion in their diagnose making. Also a single human expert may make different
diagnoses in different screening times due to human factors, such as tiredness or
sickness. A computer-based screening system does not have to be perfect to be used in
screening. It is better to use a computer-based screening system for classifying only
clearly normal funduses as normal, whereas abnormal and obscure funduses are
delivered to a human expert for further classification. However, the computer-based
screening system reduces the workload of the human expert, since in the screening
most of the funduses are normal, and only a few funduses have retinopathy. \\
Diabetes caused pathological alterations in the retinal blood vessel network that need
to detection and treatment as soon as possible in order to prevent DR.
For improving the evaluation of the retina condition image processing techniques are
required to investigate appropriate data about changes in retinal vasculature.
We divided our images into a training set with 40 images and test set with 220 images.
The training set was used for developing the algorithms and the test set only for
testing the algorithms.
Our results for fovea detection are acceptable. The results proved that it is possible to
use algorithms for assisting an ophthalmologist to segment fundus images into normal
parts and lesions, and thus support the ophthalmologist in his or her decision making.
The algorithms detect regions where the image quality is inadequate, and thus it is
possible to show to the ophthalmologist what regions are left unprocessed.
In the screening it would be important that the computer-based system has very high
sensitivity.\\
The quality of our segmentation
depends on some parameters such as the length of our window (n), measure of step,
line validation thresholding, etc. determining of these parameter appropriately has
some advantages in our processing likes:\\
Accurate detection of retinal vessels location, Determination of some parameters like width and length of vessels, and Even determining of location of vessel bifurcation which can be assist to
clinician for analyzing image in later by registration scheme.\\
Our algorithm has some important characteristics in detection of vascular structure in
retinal images that include:\\
1. This algorithm is able to determine location, width and angle of vessels.\\
2. The algorithm is robust to noise.\\
3. Because of combination of two methods vessel segmentaion methods with the multi-overlapping window the performance of algorithm in detection of thick and
even thin vessels is acceptable.\\

% or


\appendices
%\section*{Acknowledgment}

%The authors would like to thank J. J. Staal and his colleagues, A. Hoover, and Fraz and his colleagues for making their databases publicly available. 

\ifCLASSOPTIONcaptionsoff
  \newpage
\fi

\end{document}